\documentclass[referee]{cjaa}           
\usepackage{graphicx}                   
\input{epsf.sty}                        
\input{psfig.sty}                       
\def \hcm {\hbox {\ifmmode $ cm$^{-2}\else cm$^{-2}$\fi}}

\def\approxgt{\mathrel{\hbox{\rlap{\lower.55ex \hbox {$\sim$}}
        \kern-.3em \raise.4ex \hbox{$>$}}}}
\def\approxlt{\mathrel{\hbox{\rlap{\lower.55ex \hbox {$\sim$}}
        \kern-.3em \raise.4ex \hbox{$<$}}}}

\begin{document}
   \title{Mass and Angular Momentum of Black Holes: An Overlooked Effect of 
General Relativity Applied to the Galactic Center Black Hole Sgr A$\sp *$
}
   \author{B. Aschenbach
      \inst{}\mailto{}
                }
   \offprints{B. Aschenbach}                   
   \institute{Max-Planck-Institut f\"ur Extraterrestrische Physik, 
      P.O. Box 1312, Garching, 85741, Germany\\
             \email{bra@mpe.mpg.de}
          }
   \date{Received~~2006 month day; accepted~~2006~~month day}
   \abstract{
I report the discovery of a new effect of General Relativity which is
important to understand very rapidly rotating (Kerr) black holes.
The orbital velocity of a test particle is no longer a monotonic function
of the orbit radius when the spin of the black hole is $>$0.9953, but
displays a local minimum-maximum structure for radii smaller
than 1.8 gravitational radii. There the
rate of change of the orbital velocity per radius unit equals the radial
epicyclic frequency and is exactly one third of the polar epicyclic
frequency, suggesting a 3:1 resonant oscillatory motion of the particle.
If associated with the most recently observed quasi-periods the mass of
the supermassive black hole Sgr A$\sp{*}$ in the centre of the our Galaxy
is determined to 3.3$\times$10$\sp 6 M\sb{\odot}$, and the spin is 
0.99616.
   }
   \authorrunning{B. Aschenbach}            
   \titlerunning{Black hole QPO's in Sgr A$\sp *$ }  
\maketitle
\section{Introduction}           
\label{sect:intro}
The classical problem of a test particle orbiting a rotating black hole has long been solved (Bardeen et al. 1972). 
Stable circular orbits exist down to a minimal orbital radius $r$, the innermost marginally stable
circular orbit.

As for Newtonian mechanics both the energy $E$ and the angular momentum $L$ of the
particle are monotonic functions of $r$ for the full range of the black hole spin parameter $a$, the
normalized angular momentum for which -1$\le a \le$ 1.
This
means that there is no obvious preference for any specific value of $r$ and $a$ for
a particle to take. This strictly monotonic
behaviour with $r$ and $a$ has generally been assumed to hold for the orbital velocity
$v\sp{(\Phi)}$
as well. But this is not the case. 

\section{The strange behavior of the orbital velocity}           
\label{sect:strange}

By a detailed numerical analysis of the
Boyer-Lindquist functions (Boyer \& Lindquist 1967), which describe the space-time of a Kerr black hole,
 I have shown that the monotonic behaviour of $v\sp{(\Phi)}$ breaks
down for ${a~>~0.9953}$, and $v\sp{(\Phi)}$ develops a minimum-maximum structure in $r$-space
(c.f. Fig. 1) (Aschenbach 2004).

\begin{figure*}
\includegraphics[width=12cm,
 bb=43 36 570 607,
angle=-90,clip]{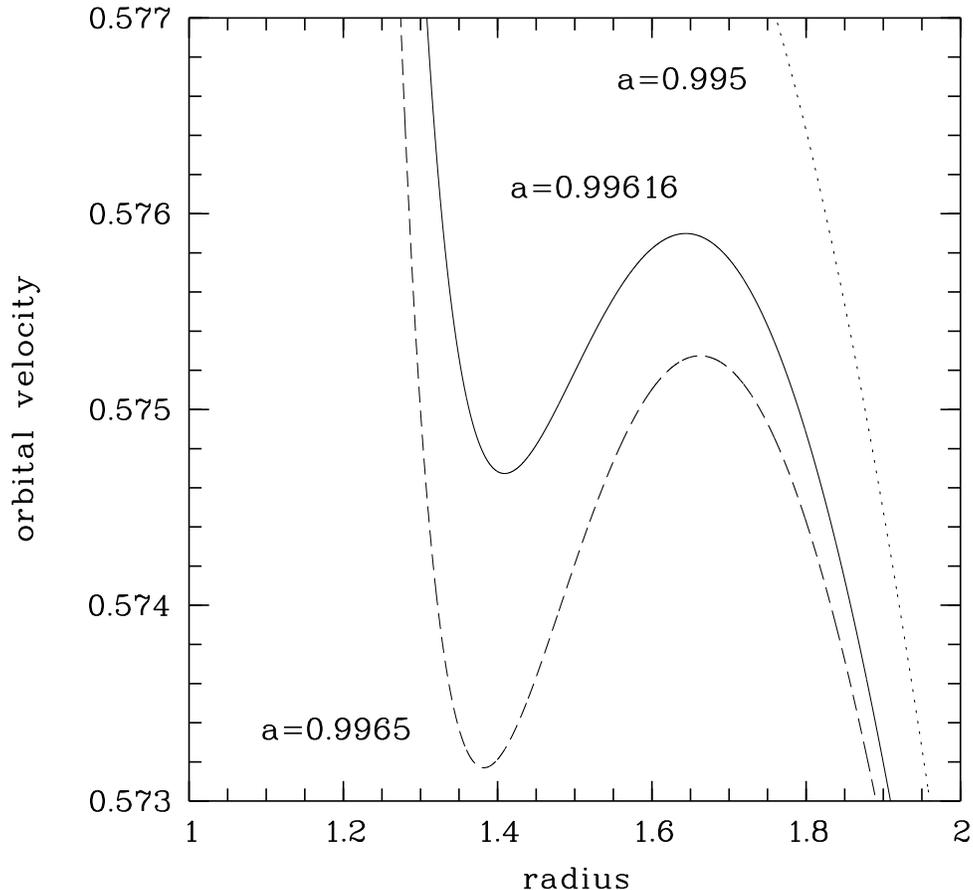}
\caption{Orbital velocity in units of $c$ vs. orbital radius $r$ in units of the
                     gravitational radius for various black hole spins $a$.}
\end{figure*}

It can be shown that the radii with $r_{min} \le r \le r_{max}$ are larger
than the corresponding radius of the innermost stable circular orbit (Aschenbach 2004). Therefore,
according to the formal Bardeen stability criterion these orbits
are stable (c.f. Fig. 2). The non-monotonic behaviour of $v\sp{(\Phi)}$ is a new effect of General Relativity which has been
overlooked so far. Meanwhile the effect has been confirmed and Stuchl\'ik et al. (2005)  have shown
that this effect occurs not only for geodesic, stable, circular orbits but also for
nongeodesic circular orbits with constant specific angular momentum (Stuchl\'ik et al. 2005). The physical relevance
and impact of this new effect are currently investigated. Variation of the topology for particles and
fluids orbiting rapidly rotating black holes is a possible consequence (Stuchl\'ik et al. 2005).

\begin{figure*}
\includegraphics[width=12cm,
 bb=75 444 470 815,angle=0,clip]{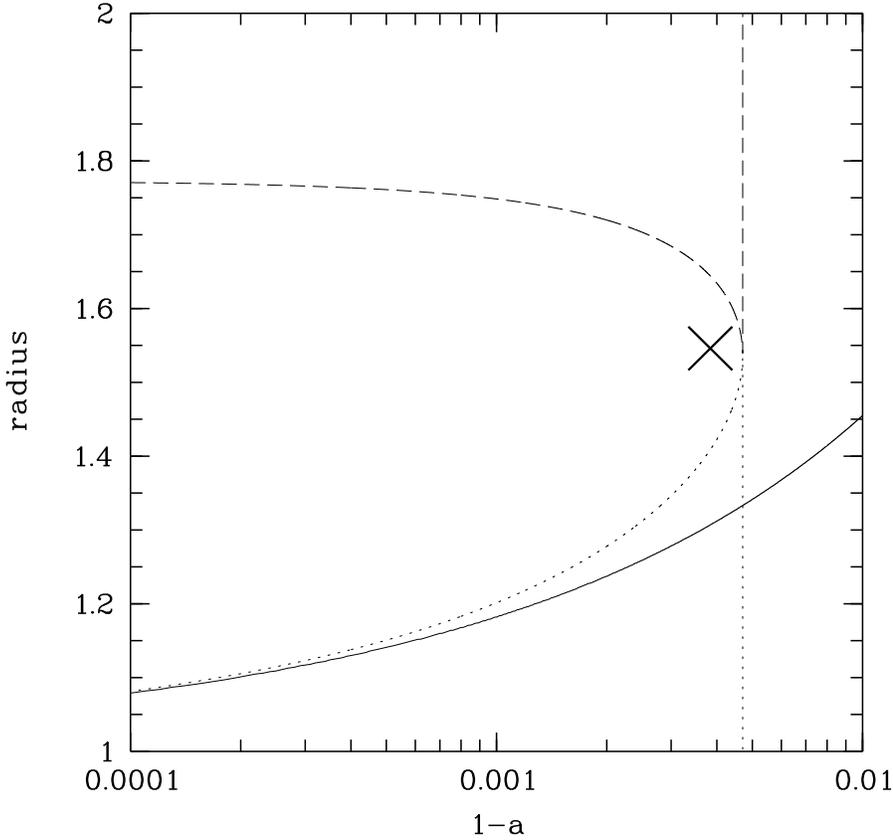}
\caption{Radii, in units of the gravitational radius, of 
         the local maximum/minimum of the orbital velocity (curved dashed/dotted line) as function
               of $1 -a$. The solid line marks the radius of the innermost stable orbit. The vertical
               dashed line represents the critical spin $a$ = 0.9953.}
\end{figure*}

According to standard
definition the orbits between $r\sb{min}$ and $r\sb{max}$ are stable but this
 region is suspicious because it is the only region
with $\partial{v\sp{(\Phi)}}/\partial r \ge 0$, which represents
a positive rate of change of the orbital velocity per
unit length of $r$ or some sort of time scale that can be compared with other typical time scales. These are
given by the orbital (Kepler) frequency, the radial epicyclic frequency $\Omega\sb{R}$ and
the vertical (polar) epicyclic frequency $\Omega\sb{V}$. These three frequencies take identical values
in Newtonian physics but all three differ from each other only in Kerr space-time for $a \ne 0$.
I therefore define a critical angular
'frequency' $\Omega\sb{c} = 2\pi {{\partial{v\sp{(\Phi)}}}\over{\partial r}}\mid\sb{max}$
at that radial position where $\partial{v\sp{(\Phi)}}/\partial r$ has the
maximum value for a given $a$.
Fig. 3 shows $\Omega\sb{c}$ in comparison with $\Omega\sb{R}$ and illustrates that  $\Omega\sb{c}$ is very
close to  $\Omega\sb{R}$ for $1 -a = 0.004$, i.e. the rate of change of the orbital velocity in radial
direction equals the epicyclic frequency again in radial direction.

\begin{figure*}
\includegraphics[width=12cm,angle=-90,clip]{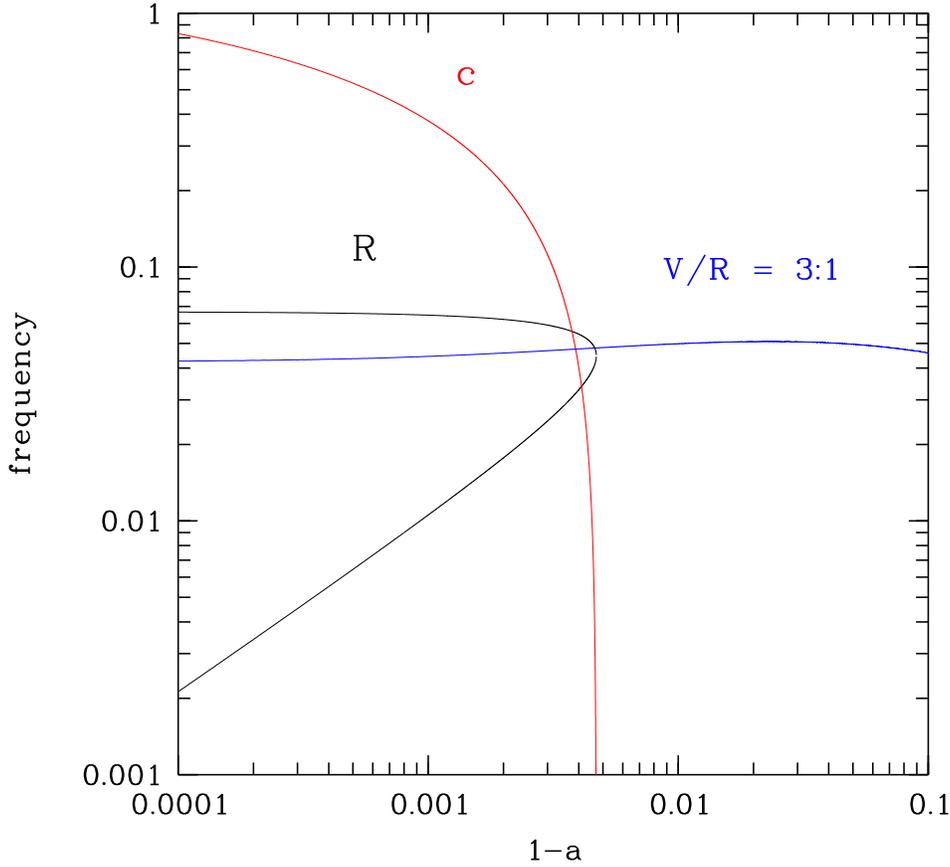}
\caption{Critical frequency (c), radial epicyclic frequency (R)
               at $r_{min}$ and $r_{max}$ (upper and lower branch of the asymmetric parabolic curve)
               and the vertical epicyclic frequency divided by a factor of three
               (V/R = 3:1) are shown; a common intersection occurs around
               $1 - a = 0.004$.}
\end{figure*}

Any pertubation at this point
in space-time may trigger and maintain radial quasi-periodic motions with $\Omega\sb{R}$.
Fig. 3 also shows that a common intersection shows up for $\Omega\sb{V} = 3 \times \Omega\sb{R}$.
A 3:1 resonance exists
between the vertical and radial epicyclic motions, and this is the lowest and only resonance
in the region defined
by $r_{min}$ and $r_{max}$. The associated spin and the radius of the orbit take unique values, which are
$a =0.99616$ and $r = 1.546$. With these parameters fixed the determination of the mass $M$ of the
black hole is just a
matter of measuring the two frequencies and confirming their 3:1 frequency ratio, and
$M = 4603.3/\nu_{up}$, with $M$ measured
in units of the sun's mass and $\nu_{up}$, the higher of the two frequencies, in Hz. 
Of course, harmonics and/or beat frequencies may
exist as well (Aschenbach 2004).

\section{Black holes in astrophysics}           
\label{sect:bh}

The existence of astrophysical black holes in the universe has been discussed for quite some time.
They have been searched in X-ray binaries with masses typical of stars, and in the cores of galaxies as
supermassive black holes with masses exceeding a few million times the mass of the sun and more.
On the stellar-mass level promising candidates have been found in low mass X-ray binaries: With a
mass significantly exceeding that of the most dense, stable object known, i.e. a neutron star,
a black hole is the only candidate for the dark component in the respective binary. About 20 such systems have
been detected so far in our Galaxy. The most promising candidate for a supermassive black hole is an
object in the centre of the Milky Way, called Sagittarius A$^*$ (Sgr A$^*$). In either case the motion
of a star, or several stars in case of Sgr A$^*$, orbiting the candidate black hole, is used for the
determination of the mass of the compact and dark object through Kepler's laws, i.e. the application of Newtonian
physics.

\par

Black holes themselves do not emit. The radiation we receive is generally from gas or dust in the vicinity of the
black hole. This matter is orbiting the black hole and may be organized in an accretion disk. Some fraction of this matter
may show oscillations and the majority of the scientists tend to associate these with Kepler frequencies.
But as shown above the epicyclic frequencies may be equally or even more
important.

\subsection{The Case of Sgr A$\sp *$}

With the dramatic improvement of the near-infrared capabilities of the large ground-based
telescopes over the last ten years and with the launch of the big X-ray observatories Chandra (NASA) and
XMM-Newton (ESA) about six years ago  the radiation of Sgr A$^*$ could be studied in great detail.
A major surprise was the discovery of huge flares both in the near-infrared and X-rays. The first report of
 a quasi-period
of about 16.8 min was reported by Genzel et al. (2003). The Fourier analysis of the X-ray flares confirmed this period
 and revealed additional quasi-periods (Aschenbach et al. 2004). The statistical signifance of the periods per observation
is not overwhelming but the fact that they have been observed in more than one independent experiment with
excessive power density favours the existence of a true signal. 
Fig. 4 shows a summary of these frequencies. 

\par

Meanwhile later measurements, both in the infrared and X-rays, have supported the presence of quasi-periods in flares. Yusef-Zadeh et al. (2006) 
have reported a period 33$\pm$2 minutes 
in HST NICMOS (1.60, 1.87, and 1.90 $\mu$m) data, 
which, within the error bars, agrees very well with the X-ray frequencies of group \#1. 

\par

From the analysis 
of an X-ray flare of fairly moderate flux observed with XMM in fall 2004 B\'elanger et al. (2005, see also Liu et al. 2006) claim a quasiperiod 
of 21.4 minutes, which has been incorporated in figure 4 as well (XMM(2)). Looking at just the four data points of 
group \#2 there seems to be a tendency 
that we observe different frequencies in those four measurements, 
although a rigorous error analysis would not exclude a single frequency for group \#2.  
Interestingly, both Genzel el al. (2003) and Liu et al. (2006) claim that the period they 
found is not constant over the observation but is decreasing 
from 22.7 minutes to 11.8 minutes (Genzel et al. 2003) and 25 minutes to 17.5 minutes (Liu et al. 2006) based on the separation of the 
flux maxima/minima. Whether this is actually a decrease of the period 
remains to be seen, given the fact that just six to nine minima/maxima with fairly low counting statistics and complex lightcurves have been observed. 
These authors advocate for Keplerian motion with decreasing period because of radial infall with a velocity of 
about 0.3\%$\times$c.   
Independent of the interpretation these results, because of their short periods - if they are actually orbital 
periods -  underline that the modulation of the flux happens within a few Schwarzschild radii of the black hole and that the Sgr A$\sp *$ 
black hole  
has a significant spin.

\begin{figure*}
\includegraphics[width=12cm,angle=-90,clip]{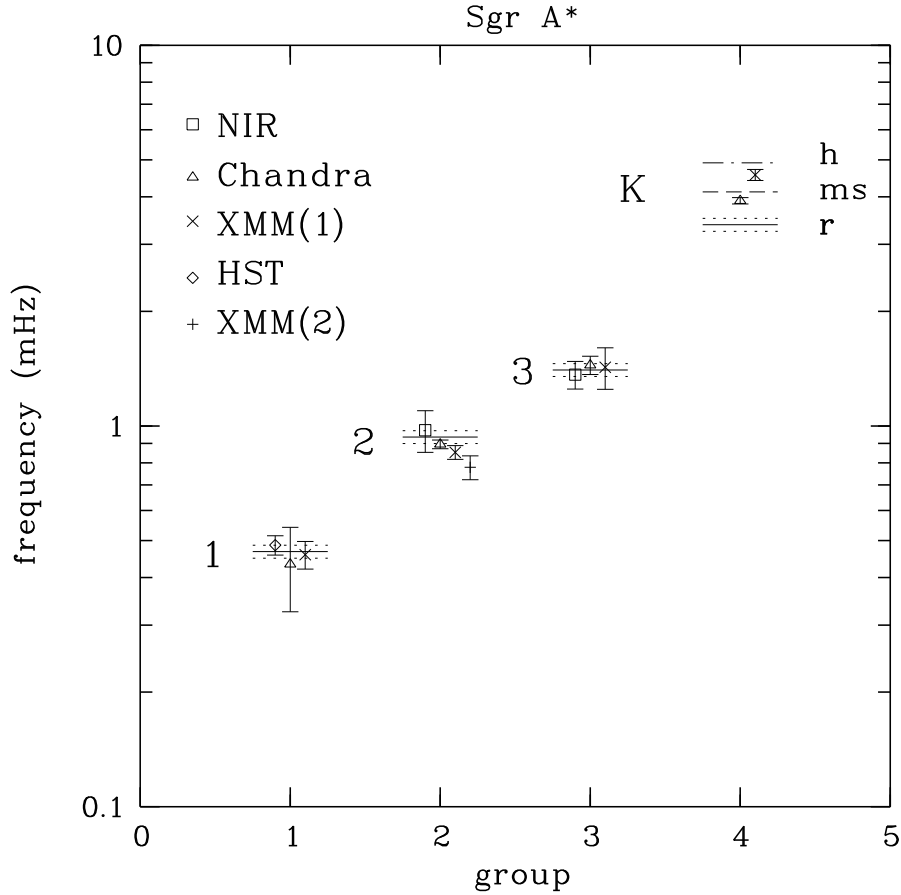}
\caption{Quasi-periodic oscillations discovered in at least two
               independent power density spectra of Sgr A$^*$, in the near-infrared from the ground (NIR, Genzel et al. 2003),
               X-rays (Chandra and XMM(1), Aschenbach et al. 2004), the infrared with the Hubble Space Telescope (HST, Yusef-Zadeh et al. 2006) 
               and XMM again (XMM(2), Liu et al. 2006; B\'elanger et al. 2005). 
               They fall into four groups; the frequencies of the first three groups come in a ratio
               of 1:2:3. The best fit to such a ratio sequence is indicated by the
               horizontal solid lines. The $\pm$1 $\sigma$ errors of the best fit
               are represented by the horizontal dashed lines.}
\end{figure*}

Basically, the frequencies can be divided into four groups. The frequencies of the first three groups appear to follow a ratio of
1:2:3, which is confirmed by a best fit.

Because of the appearence of a 3:1 resonance I have associated the highest of the three frequencies
with the vertical epicyclic frequency and the lowest of the three frequencies with the radial
epicyclic frequency, and I assume that the appearence of these frequencies is due to the new effect
of General Relativity described above. In this way the spin $a$ and the orbit radius $r$ are fixed and the mass of the
black hole is set by the observed frequencies. For the numerical relation between mass and frequency one may take either 
the vertical or the radial frequency. 
Determined in this way
it turns out that the mass of the Sgr A$^*$ black hole
is $(3.28 \pm 0.13)\times$10$\sp 6$ times the mass of our sun. I stress that this mass estimate follows from
a straight inverse relation between mass and frequency and does not depend on any other observables like the distance to
the black hole. Fig. 5 shows a comparison with the most recent estimates of the Sgr A$^*$ mass which has been derived
from more than a decade long dynamical measurements of the orbits of the S-stars around Sgr A$^*$
 (Eisenhauer et al. 2005, Ghez et al. 2005). These results depend
on the distance to Sgr A$^*$, and the respective data shown are normalized to the same best estimate of the
distance of 7.62 kpc (Eisenhauer et al. 2005).
The agreement is more than satisfactory and there is the chance to lower the uncertainty for the QPO measured mass significantly 
by improving the frequency measurements.

\begin{figure*}
\includegraphics[width=12cm,angle=-90,clip]{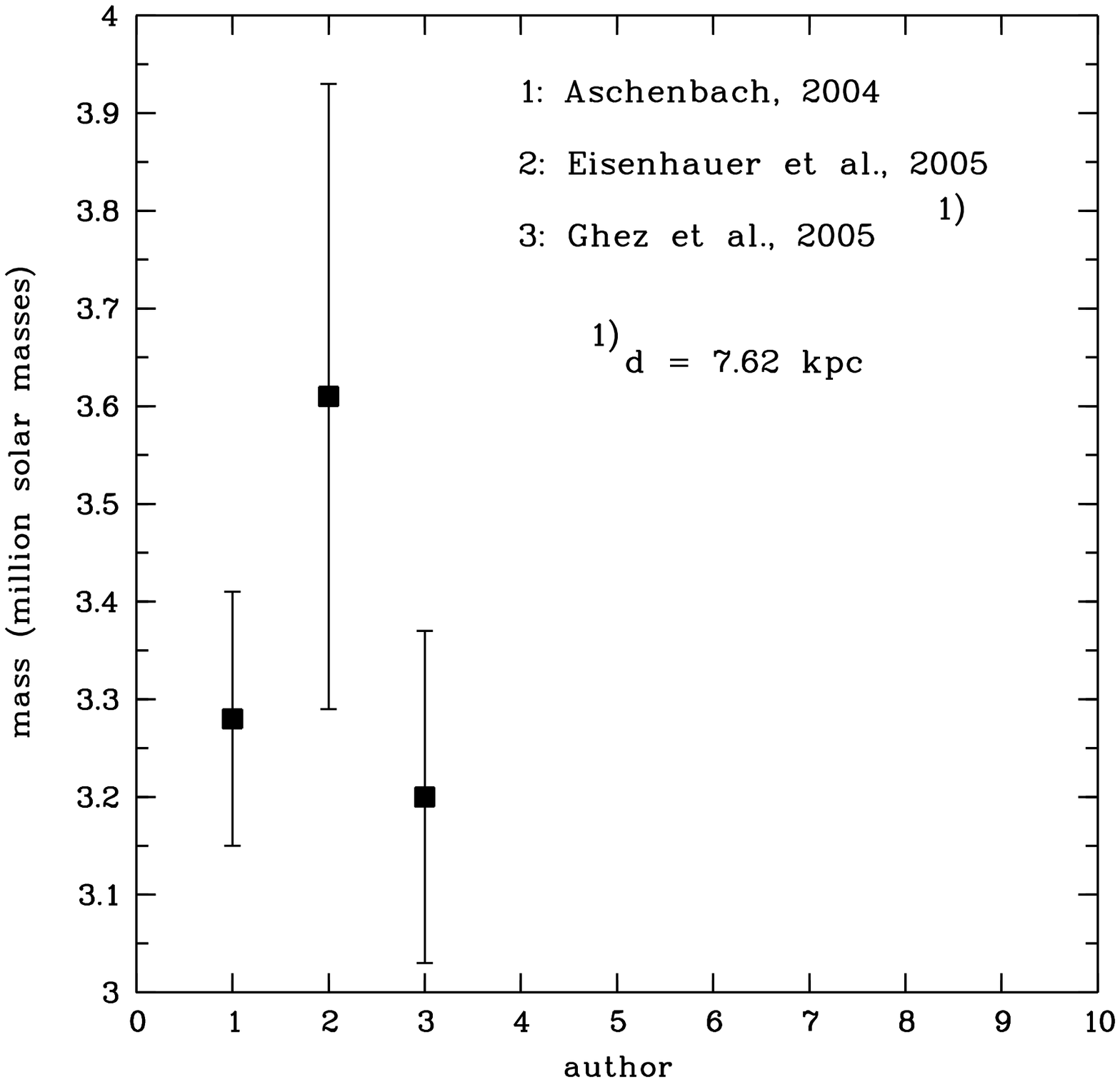}
\caption{Mass of the black hole Sgr A$^*$ in the centre  of our Galaxy.
               Data 1: mass determined by epicyclic frequencies (Aschenbach 2004), data 2 and 3: mass determined
               dynamically by measurements of stellar orbits (Eisenhauer et al. 2005, Ghez et al. 2005);
               these data have been normalized to the best estimate
               of the distance  of d = 7.62 kpc (Eisenhauer et al. 2005).}
\end{figure*}

As a side product I mention that these QPOs arise in a region very close to the black hole at $r= 1.546$ or 0.773
Schwarzschild radii, which is an excellent test bed for the effects of strong gravity. For example, the relative difference
of the orbital velocity treated by either General Relativity or Newtonian mechanics is more than 65\%.

Finally I draw the attention to the frequencies of group \#4 in Fig. 4. As spin and mass of the black hole are
fixed the Kepler frequency can be calculated as a function of orbital radius. The largest radius considered is the orbit at which
the epicyclic oscillations are launched (label r), furtherin I consider the last marginally stable orbit (label ms) and at last the radius
of the event horizon (h). The two frequencies observed with Chandra and with XMM-Newton, if confirmed by
future measurements, demonstrate that we can in principle see quasi-periodic radiation from below the last stable orbit almost down
to the event horizon. Of course these events are likely to be transient and the frequency can vary from one
observation to the other between 3.3 and almost 5 mHz depending on the orbital radius. The Kepler frequency may not be strictly periodic,
which makes the detection and confirmation difficult and instruments with very large collecting areas exceeding the capabilities of
the present generation of telescopes are required.

\section{Conclusions}
\label{sect:conclusion}
A new effect concerning the physics of fast spinning Kerr black holes is presented. The relevant range of black hole 
spin and orbital radii of test particles coincides with regions where a 3:1 resonance between vertical and radial 
epicyclic frequencies occurs, if the spin is greater than a = 0.9953 and the radius is less than 1.8 gravitational radii. 
The new effect of a decreasing orbital velocity with decreasing orbital radius may trigger the epicyclic oscillations. 
Infrared and X-ray observations have indicated the presence of such quasi-periods in flares from Sgr A$\sp *$, and there 
is growing evidence that these QPOs are real despite their low statistical significance per observation. If the QPO's, which come 
in a 1:2:3 ratio, are taken as epicyclic frequencies the mass of the Sgr A$\sp *$
black hole is uniquely determined to 
3.3$\times$10$\sp 6 M\sb{\odot}$ with an error as big as the frequency measurement error. 
Independent of the association of frequency with oscillation mode the observed QPOs demonstrate that the 
light modulation of the flare emission occurs in the lowest few Schwarzschild radii of the black hole. 
For the unambiguous determination of the spin of the black hole it will be essential to cover the shortest 
periods possible. The NIR flare reported by Genzel et al. (2003) shows a separation between two 
relative maximum flux values as short as 
 11.8 minutes and the XMM(1) flare goes down to a period of about 219 sec. Of course this needs to be 
confirmed, but if it is, the observation window of black holes might not only cover the region of the last stable orbit but 
stretches down close to the event horizon.

\bigskip 
\noindent
{\b DISCUSSION}
\bigskip

\noindent
{\b DIDIER BARRET:} Could you please explain a bit more the reasons to identify the maximum observed frequency
with the vertical epicyclic frequency.
\bigskip 

\noindent 
{ASCHENBACH:} For a Kerr BH there are three epicyclic frequencies following 
the three axes, i.e. orbital (Kepler), polar (vertical) and 
radial. For any fixed orbital radius and fixed BH spin the Kepler frequency is always greater than the 
vertical epicyclic frequency, which is always greater than the radial epicyclic frequency. The model I propose 
assumes a resonance between the vertical and radial frequencies with a ratio of 3:1. If that ratio is observed 
clearly the higher frequency of the two observed is to be associated with the vertical epicyclic frequency. 
The Kepler frequency at that orbit radius and that spin is even higher but is nowhere in the orbit in 
resonance with any one of the other two frequencies. 
\bigskip

\noindent
{\b DIDIER BARRET:} How sensitive are your results on the mass with respect to the 
identification of the QPO frequency?
\bigskip

\noindent
{ASCHENBACH:} In this respect the association of the QPO frequency with a particular oscillation type is absolutely 
essential. In the relevant equations there are the following four unknowns: the BH mass, the BH spin, the 
orbital radius at which the oscillation occurs and the oscillation type. To solve that problem unambigously 
we have to have knowledge or constraints of at least three of the four unknowns. Usually observers 
assume Kepler orbital motion at the innermost stable orbit to be responsible for the observed frequency
which would result in a BH mass BH spin relation. With two frequencies observed the problem is somewhat 
better defined. The important issue in my work is the discovery of the strange behaviour 
of the orbital velocity for high values of the spin at low orbits and if this is coupled to the 
resonant behaviour of the epicyclic frequencies, mass and spin are uniquely determined and the 
mass is as precise as the frequency measurements are. A resonant behaviour involving the Kepler 
frequency is definitely excluded for the high spin low orbit case because the resulting BH mass is 
by far too low compared with the dynamical mass. 
\bigskip
    
\noindent
{\b DANIELE FARGION:} Does General Relativity explain jets? 
\bigskip

\noindent
{ASCHENBACH:} I have no straight answer to that, but I think one should not rule out such a 
possibility (c.f. Aschenbach 2004). If the vertical epicyclic oscillations exist and are created close to the inner 
edge of the accretion disk, why shouldn't mass and energy be expelled 
by them perpendicular to the accretion disk. This definitely needs further study.
\bigskip

\noindent
{\b DANIELE FARGION:} Did you consider jets drag on your frequency? 
\bigskip

\noindent
{ASCHENBACH:} No, I did not. 
\bigskip

\noindent
{\b WOLFGANG KUNDT:} Congratulations on your findings. But when discussing the alternative 
possibilities you did not mention the Supermassive Magnetized Disk (SMD) / burning disk model 
(e.g. W. Kundt in Astrophysics and Space Science, 62, 335 (1979)).  
\bigskip

\noindent
{ASCHENBACH:} In the introduction of this talk I have tried to summarize the arguments that  
have led people to state "the most convincing case for the existence of an astrophysical black hole is 
Sgr A$\sp *$". This statement has been made by various scientists, for instance, at the 
Joint Astronomy Conference 'Growing Black Holes: Accretion in a Cosmological Context' held at Garching 
in June 2004. The general consensus seems to be that we are still lacking 'the proof' for 
Sgr A$\sp *$ being a black hole. Therefore other models are still something to persue. 
In my talk I used the physical properties expected for black holes and used them to understand 
the QPOs, and as far as I am concerned I don't see any contradiction, which does not tell that they are 
eventually nothing more than a viable model.
\bigskip 

\noindent
{\b NANDA REA:} Do you expect any spectral variability during these oscillations?   
\bigskip

\noindent
{ASCHENBACH:}
If strong gravity effects produce the light variations as Marek Abramowicz and others
have recently suggested I would expect some spectral variations.
\bigskip

\noindent
{\b NANDA REA:} Did you check for spectral variations doing a phase resolved spectroscopy during the
 XMM and Chandra flares?  
\bigskip

\noindent
{ASCHENBACH:} Yes we did that, but we did not find any significant 
variations. But I stress that the relative flux of the modulated signal is very
low. X-ray telescopes with much larger collecting area are needed.
\bigskip
       
\label{lastpage}
\end{document}